\documentclass[sigconf]{acmart}
\usepackage{multirow}
\usepackage{multicol}
\usepackage{graphicx}
\usepackage{subcaption}

\newcommand{\reporank}{\texttt{GitRank}}

\acmConference[MSR 2022]
  {MSR '22: Proceedings of the 19th International Conference on Mining Software Repositories}
  {May 23–24, 2022}{Pittsburgh, PA, USA}

\author{Niranjan Hasabnis}
\affiliation{
  \institution{Intel Labs}
  \streetaddress{3600 Juliette Ln}
  \city{Santa Clara}
  \state{California}
  \country{USA}
}

\email{niranjan.hasabnis@intel.com}

\title{{\reporank}: A Framework to Rank GitHub Repositories}

\copyrightyear{2022} 
\acmYear{2022} 
\setcopyright{acmlicensed}\acmConference[MSR '22]{19th International Conference on Mining Software Repositories}{May 23--24, 2022}{Pittsburgh, PA, USA}
\acmBooktitle{19th International Conference on Mining Software Repositories (MSR '22), May 23--24, 2022, Pittsburgh, PA, USA}
\acmPrice{15.00}
\acmDOI{10.1145/3524842.3528519}
\acmISBN{978-1-4503-9303-4/22/05}

\begin{document}
\begin{abstract}

Open-source repositories provide wealth of information and are increasingly
being used to build artificial intelligence (AI) based systems to solve
problems in software engineering. Open-source repositories could be of varying
quality levels, and bad-quality repositories could degrade
performance of these systems. Evaluating quality of open-source repositories,
which is not available directly on code hosting sites such as GitHub, is thus
important. In this hackathon, we utilize known code quality measures and
GrimoireLab toolkit to implement a framework, named {\reporank}, to rank
open-source repositories on three different criteria. We discuss our findings
and preliminary evaluation in this hackathon report.

\end{abstract}
\maketitle

\section{Introduction}

Open-source code repositories, such as those hosted on popular hosting sites
like GitHub, provide wealth of information about software projects such as team
dynamics, project schedule and deliverables, among
others~\cite{banker:1998:software, bird:2011:donttouch, hudson:2018:github_stars,
munaiah:2017:curating}. Such information is routinely used by researchers
working on problems in the field of software engineering. More recently,
researchers working at the intersection of machine learning (or AI) and software
engineering are also using such information to build systems that automatically
learn to solve problems from data~\cite{chen:2021:codex, hasabnis:2016:lisc,
hasabnis:2016:eissec, hasabnis:2020:controlflag, luan:2019:aroma,
roziere:2020:transcoder, Yasunaga:2020:drrepair, ye:2020:misim}.

Unfortunately, deriving insights using open-source code repositories could be
wrong or misleading because popular hosting sites such as GitHub typically do
not indicate quality of code repositories. This could partly be because quality
of a repository could be subjective. Nonetheless, measuring code quality a
well-researched problem in the field of software engineering, and several,
popular source code metrics such as Halstead volume~\cite{halstead}, cyclomatic
complexity~\cite{mccabe:1976:cc}, maintainability index~\cite{oman:1992:mi},
etc., exists~\cite{Jarczyk:2014, ludwig:2017, ludwig:2019:cbr}.

Although, a number of frameworks and source code metrics to measure
repository/code quality exist~\cite{Chatzidimitriou:2018:npm_miner, iso25000,
Jarczyk:2014, samoladas:2008:sqooss, spinellis:2009:sqooss1,
stamelos:2002:codequality}, we could not find publicly-available, open-source
implementation of a framework that ranks code repositories on quality. That is
why we decided to build such a framework, named
{\reporank}\footnote{Available at \url{https://github.com/nirhasabnis/gitrank}}, in this
hackathon.  Additionally, GrimoireLab toolkit~\cite{duenas:2021:grimoirelab, grimoirelab} that is
available as a part of this hackathon incentivized such an implementation as it
makes retrieving some of the source code metrics easy. We, nonetheless, faced
several challenges, which we discuss in this report.

\section{{\reporank}: Design and Implementation}
\label{section:framework}

\begin{figure*}[!htbp]
\centering
\includegraphics[width=0.95\textwidth]{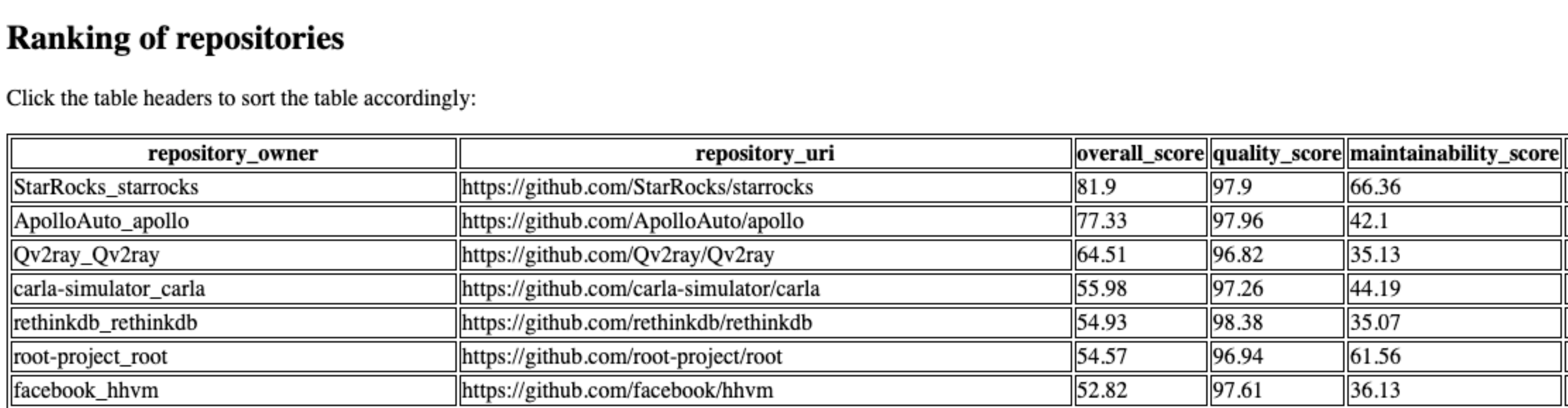}
\caption{An HTML output of {\reporank}}
\label{fig:html_output}
\Description{An HTML output of {\reporank}}
\vspace{-0.1in}
\end{figure*}

We now describe our framework to rank open-source repositories based on quality,
maintainability, and popularity. In addition to source code quality metrics,
{\reporank} also considers popularity and maintainability metrics (similar to
NPM registry~\cite{npmregistry}).
{\reporank} currently supports ranking of repositories that use C/C++ as their
primary language.  Nonetheless, the design of our framework is easy to extend to
other languages as well. {\reporank} is implemented in ~700 lines of Python code
and can be broken down into two phases.

\paragraph{\textbf{Phase 1: Obtaining values of metrics for every repository}}

Given a set of open-source repositories to be ranked, in the first phase, we
obtain values of quality, maintainability, and popularity metrics of every
repository individually.  Table~\ref{table:measures} contains the set of
measures that we consider.  We use a subset of the typical
code measures used in software engineering that are applicable to open-source
repositories and are common across a set of languages.

\begin{table}
\begin{center}
\begin{footnotesize}
\begin{tabular}{l|l}
\hline \hline
\textbf{Category} & \textbf{Description of the measure} \\
\hline \hline
\multirow{5}{*}{Quality} & Average Cyclomatic complexity of a repository \\
 & Number of style errors per LoC \\
 & Number of security errors of low severity per SLoC \\
 & Number of security errors of medium severity per SLoC \\
 & Number of security errors of high severity per SLoC \\
\hline
\multirow{6}{*}{Maintainability} & Average maintainability index of a repository \\
 & Number of closed issues and pull requests over last 2 years \\
 & Number of closed issues and pull requests over last 1 year \\
 & Number of closed issues and pull requests over last 6 months \\
 & Number of closed issues and pull requests over last 1 month \\
 & Number of commits per day \\
\hline
\multirow{3}{*}{Popularity} & Number of subscribers per day \\
 & Number of stargazers per day \\
 & Number of forks per day \\
\hline
\end{tabular}
\end{footnotesize}
\end{center}
\caption{Measures used in {\reporank}}
\label{table:measures}
\vspace{-0.36in}
\end{table}


\paragraph{Cyclomatic complexity.} Cyclomatic complexity of a source code is a well-known metric of the
structural code complexity~\cite{mccabe:1976:cc}.  We use Python APIs of
\texttt{lizard} project~\cite{lizard} to obtain function-level cyclomatic
complexity and then average it over a repository to obtain repository-level
cyclomatic complexity\footnote{We ended up using \texttt{lizard} API directly as CoCom
backend of Graal component in GrimoireLab~\cite{grimoirelab} offered commit-specific code
complexity instead of function-level or repository-level complexity.}.

\paragraph{Style errors.} We consider code formatting as a code quality metric
and use \texttt{cpplint} project~\cite{cpplint} to obtain the number of style errors (of
the highest severity level)\footnote{Graal's CoQua backend does
not support obtaining this information for C/C++.}. We divide the total
number of errors by the lines of source code
(SLoC)~\cite{Nguyen:2007:sloc} to obtain the error density. Static analysis
tools such as \texttt{cpplint} are known to suffer from false
positives~\cite{panichella:2015}, and
although, {\reporank} does not account for them right now, we believe that they
could be accounted by adjusting the weight assigned to style errors in the
overall quality score.

\paragraph{Security errors.} We also consider security errors to determine code quality.
Security issues,
however, could be of varying severity levels. In our framework, we consider
three severity levels: low, medium, and high. We used \texttt{FlawFinder}
tool~\cite{flawfinder} to obtain the security errors in C/C++
programs\footnote{We used a standalone tool to obtain this information as we
found that Graal's CoVuln backend did not support our specific case.} and divide
the total number of errors reported for every level by the lines of source code
(SLoC) to obtain the level-specific error density.

\paragraph{Maintainability index.} Maintainability index is a well-known metric that incorporates
a number of traditional source code metrics into a single number that indicates
relative ease of maintaining source code~\cite{oman:1992:mi}. We use the
modified formula of MI~\cite{welker:1997:modified_mi} to obtain MI for
individual modules from a repository\footnote{We extend \texttt{lizard} project with a
plugin to calculate Halstead volume in MI.}. MI for a repository is then obtained by
averaging over MI for individual modules.

%
%

\paragraph{Closed issues.} We use a combination of GitHub's REST APIs and
GrimoireLab's Perceval~\cite{duenas:2018:perceval}
to obtain the number of closed issues over a period of time (last 2
years, last 1 year, last 6 months, and last 1 month, with increasing importance
in that order) from the date of evaluation
to determine maintenance activity of a repository.
This is because we want to value current maintenance activity more.


\paragraph{Stars, watches, and forks.} Existing approaches consider GitHub
stars, subscriptions, and forks as popularity
metrics~\cite{hudson:2018:github_stars}. We follow the same terminology. We use
a combination of GitHub's REST APIs~\cite{github_rest_api} and GrimoireLab's
Perceval to obtain this information.


\paragraph{\textbf{Phase 2: Obtaining quality, popularity, and maintainability score of
repositories}}

In the second phase, we compare metrics across repositories to obtain overall
score for every repository.  Before we calculate scores for all the
repositories, we first normalize values of all the measures to the range of 0\%
(lowest) to 100\% (highest). This is performed by obtaining the lowest and the
highest value of every measure and computing the position of a value within the
range of the lowest and the highest value.
%

Normalized values of the measures are then used to compute the quality score and
the popularity score as a weighted average, with equal weights for all the
measures. For the maintainability score, we use different weights for the
measures based on their significance\footnote{We, however, envision that the weights
could be adjusted by the {\reporank} user.}. We then compute overall score of a
repository as a mean of all three scores.  All of these scores are also
normalized to the range of 0\% (lowest) to 100\% (highest).

\section{Evaluation}

%
%

{\reporank} accepts a list of GitHub repositories (their URLs) as
input and generates their ranked list as output (in CSV and HTML format).
In our preliminary evaluation of {\reporank}, we ranked
randomly-selected 500 GitHub repositories that use C++ as primary
language. We dropped 12 repositories out of 500 as they had some
peculiarities (such as unusual extensions of C++ files) because of which the tools
used by {\reporank} could not obtain their metrics.

\paragraph{\textbf{Performance.}} To support a large number of repositories, we
have implemented phase 1 of {\reporank} to be highly-parallel (e.g., multiple
machines and/or CPU cores).
In our experiment, we divided the list of 500 repositories into 5 separate lists
of 100 repositories each and processed them separately on 5 different
machines. Phase 1 of {\reporank} took ~1 hr and
~25 KB of memory, while phase 2, implemented in a serial manner, took ~0.1 sec
and 1.5 KB of memory.

\paragraph{\textbf{Output.}} Figure~\ref{fig:html_output} shows the HTML-formatted ranked
list of repositories in our experiment. The report also contains next level of
details such as quality, maintainability, and popularity scores.
Further levels of details (containing values of individual metrics) can be
obtained with additional verbosity levels.

\section{Conclusion}

In this report, we described our efforts to build {\reporank}, a framework to
rank open-source repositories using known source code metrics. So far, {\reporank}
supports repositories using C/C++ as their primary languages and uses a
combination of GrimoireLab toolkit and publicly-available tools for the
implementation. As per our evaluation, {\reporank} is highly-parallel and
performs reasonably well on commodity CPUs. We, however, did not
explore applications of {\reporank} in this hackathon and consider this as a
separate project that we plan to explore in the future.

\bibliography{main}
\bibliographystyle{plain}
\end{document}